\documentstyle[preprint,epsf,aps]{revtex} 
\input epsf
\newcommand{\be}{\begin{equation}}
\newcommand{\ee}{\end{equation}}
\newcommand{\bea}{\begin{eqnarray}}
\newcommand{\eea}{\end{eqnarray}}
\def\tr{\mathop{\rm Tr}\nolimits}
\def\mp{M_{Pl}}

\begin{document}

\title{Quintessence and the Rest of the World\footnote{Subtitle
``Suppressing Long-Range Interactions'' added in journal.}}	
\author{Sean M.  Carroll}
\address{Institute for Theoretical Physics, University of
California,
Santa Barbara, California 93106, USA\\ 
Email: {\tt carroll@itp.ucsb.edu}}

\maketitle

\begin{abstract}
A nearly-massless, slowly-rolling scalar field $\phi$ may provide most
of the energy density of the current universe.  One potential
difficulty with this idea is that couplings to ordinary matter, even
if suppressed by the Planck scale, should lead to observable
long-range forces and time dependence of the constants of nature.  I
explore the possibility that an approximate global symmetry serves to
suppress such couplings even further.  Such a symmetry would allow a
coupling of $\phi$ to the pseudoscalar $F_{\mu\nu}\widetilde
F^{\mu\nu}$ of electromagnetism, which would rotate the polarization
state of radiation from distant sources.  This effect is fairly well
constrained, but it is conceivable that future improvements could lead
to a detection of a cosmological scalar field.

\end{abstract}

\vfill
\noindent NSF-ITP/98-063\\ astro-ph/9806099

\eject

{\it And yet, to me, what is this quintessence of dust?} 
\begin{flushright} -- {\sl Hamlet}, Act II, Scene ii. \end{flushright}

Recently a number of pieces of evidence, especially
studies of the Hubble diagram for Type Ia supernovae \cite{supernova},
have lent support to the idea that the universe is dominated
by a smooth component with an effective negative pressure,
leading to an accelerating expansion.
While the most straightforward candidate for such a component
is the cosmological constant \cite{cpt}, a plausible
alternative is dynamical vacuum energy, or ``quintessence'' \cite{q,pgbs}. 

A number of models for quintessence have been put forward,
the most popular of which invoke a scalar field in a very
shallow potential, which until recently was overdamped in its 
evolution by the expansion of the universe.  For generic potentials the
requisite shallowness implies that excitations of the field
are nearly massless, $m_\phi \equiv \sqrt{V''(\phi)/2}
\leq H_0 \sim 10^{-33}$~eV.  To provide the necessary energy density,
the present value of the potential must be approximately the
closure density of the universe, $V(\phi_0)\sim (10^{-3}
{\rm ~eV})^4$, so the field itself will typically be
of order
$\phi_0 \sim 10^{18} {\rm ~GeV} \sim \mp = (8\pi G)^{-1/2}$, where
$\mp$ is the reduced Planck mass.  (These estimates can be dramatically
altered in models with more complicated dynamics \cite{ac}.)

The exchange of very light fields gives rise to forces of very
long range, so it is interesting to consider the direct interaction
of the quintessence field $\phi$ to ordinary matter.
Although it is traditional to neglect (or set to zero) the
couplings of this light scalar to the standard
model, we expect that our low-energy world
is described by an effective field theory obtained by integrating
out degrees of freedom with momenta larger than some mass scale
$M$, in which case it is appropriate to include non-renormalizable
interactions suppressed by appropriate inverse powers of $M$.
For example, $\phi$ can couple to standard-model fields via
interactions of the form
\be
  \beta_i {{\phi}\over M}{\cal L}_i\ ,
  \label{inter}
\ee
where $\beta_i$ is a dimensionless coupling and ${\cal L}_i$ 
is any gauge-invariant dimension-four operator, such as
$F_{\mu\nu}F^{\mu\nu}$ or $i\bar\psi \gamma^\mu D_\mu \psi$. 
In the absence of detailed knowledge about the structure of the theory
at high energies, the couplings $\beta_i$ are expected to be of
order unity.  

The mass parameter $M$, meanwhile, represents the 
energy scale characterizing the phenomena
which we have integrated out to obtain the low-energy
description.  We cannot specify it with precision, 
but it should not be higher than the scale where quantum
gravity becomes relevant -- not only may there be new particles
at this energy, but exotic effects such as wormholes and virtual
black holes become relevant \cite{qg}.  With this in mind,
the limits in this paper will be quoted in terms of
the reduced Planck mass $\mp\sim 10^{18}$~GeV, but cases could
be made for values as high as the traditional Planck mass
$G^{-1/2}\sim 10^{19}$~GeV or as low as the unification scale
$M_{\rm unif}\sim 10^{16}$~GeV (for example, in the
phenomenologically attractive regime of M-theory compactified
on an interval \cite{mtheory}).

The scalar force mediated by $\phi$ 
will not obey the Equivalence Principle
(which is only compatible with forces mediated by spin-two
fields), and hence is constrained by E\"otv\"os-type experiments.
Su {\it et al.} \cite{su} have found that the differential acceleration 
of various test bodies, in the
direction of the Sun, is less than $10^{-12}$ times the strength
of gravity.  Such limits can be translated into constraints on
the dimensionless couplings $\beta_i$; for example, we may calculate 
the charge on a test body due to a coupling $\beta_{G^2}(\phi/M)
\tr(G_{\mu\nu}G^{\mu\nu})$, where $G_{\mu\nu}$ is the field strength 
tensor for QCD ({\it cf.} \cite{pcw}).  Although it is difficult to 
compute QCD matrix elements to high precision, the Su {\it et al.} 
results can be used to place a conservative upper limit
\be
  |\beta_{G^2}| \leq  10^{-4} \left({M\over{\mp}}
  \right)\ .
\ee
Similar considerations constrain other couplings, although
typically not quite as well (see {\it e.g.} \cite{damour}).

A related phenomenon is the time variation of the constants
of nature.  For the dynamical nature of $\phi$ to be relevant
today, we expect a change in $\phi$ of order $\mp$ over 
cosmological timescales $t_0 \sim H_0^{-1}$.  
In that case, a coupling such as
$\beta_{F^2}(\phi/M)F_{\mu\nu}F^{\mu\nu}$ will lead to evolution
of the fine-structure constant $\alpha$.  Various observations
constrain such variation.  For example, isotopic abundances in the Oklo
natural reactor imply that $|\dot\alpha/\alpha| < 10^{-15} {\rm ~yr}^{-1}$
over the past two billion years \cite{oklo}; this leads to the limit
\be
  |\beta_{F^2}| \leq 10^{-6} \left({{M H_0}\over 
  {\langle\dot \phi\rangle}}\right)\ ,
  \label{alphalimit}
\ee
where $\langle\dot \phi\rangle$ is the average rate of change of $\phi$ 
in the last two billion years.  (There has also been a claimed detection
of a difference between the fine structure constant today and at
a redshift $z\geq 1$ \cite{alphadot}; given the preliminary nature
of the claimed detection, it is safest to rely on the limit 
(\ref{alphalimit}).)  Again, changes in other couplings are also
constrained.

There is clearly good evidence against the existence of a
nearly-massless scalar field coupled to the standard model
via nonrenormalizable interactions with strength of order $1/\mp$.
It would be premature, however, to conclude that the idea of
quintessence is ruled out, as we may consider imposing symmetries
which prevent the couplings considered thus far.  An exact continuous
symmetry of the form $\phi \rightarrow \phi + {\it const}$ is 
clearly not appropriate, as it would not allow for a nontrivial
potential $V(\phi)$.  An alternative possibility is a discrete
symmetry, for example of the form $\phi \rightarrow -\phi$; this
would forbid terms linear in $\phi$, and could arise from spontaneously 
broken gauge symmetries \cite{kw}.
However, in the case at hand discrete symmetries appear to be
ineffective, as they themselves are spontaneously broken.  (The 
field $\phi$ is expected to be displaced from the fixed point of the 
symmetry, so an effective linear term will be unsuppressed.)

We are therefore left with the possibility of approximate global
symmetries of the form $\phi \rightarrow \phi + {\it const}$.  Indeed,
such symmetries are invoked in pseudo-Goldstone boson models of
quintessence \cite{pgbs}, as an explanation for the naturalness of the
small mass $m_\phi$: in the limit as the symmetry is exact, this mass
goes to zero.  This same effect could explain the small values of the
dimensionless couplings $\beta_i$.  In this sense, the PGB models are
more likely than those based (for example) on moduli fields; in the
latter set of theories, the scalar field represents a flat direction
which typically does not generate any symmetry, with a potential
generated solely by nonperturbative effects.  There is no apparent
reason for the $\beta_i$'s to be small in such models.

An objection to this scenario is that quantum-gravity effects
do not respect global symmetries.  It is known, for example,
that there are no unbroken global symmetries in string theory
\cite{noglo}.  Furthermore, the induced interactions mentioned
above from wormholes and virtual black holes are constrained
solely by gauge symmetries \cite{qg}.
These symmetry-breaking effects have
been suggested as problems for axion and texture theories
\cite{axtex}. 

Nevertheless, although our current understanding of quantum gravity 
and string theory leads us to believe that global symmetries are
generically violated, it is insufficient to say with confidence 
that the resulting violations are in some sense large (in our
context, that the parameters $\beta_i$ are of order unity).
For example, although string theory has no exact global symmetries,
it does have axion-like fields with an approximate Peccei-Quinn
symmetry.  
It may also be the case that pure quantum gravity effects are
nonperturbative and suppressed by $e^{-S}$, where the action
$S$ can be large in specific models. Kallosh {\it et al.}
investigated this possibility in the context of axions in the
presence of wormholes \cite{klls}.  They found that the action 
was sensitively dependent on the structure of spacetime on
small scales, and there could be sufficient
suppression of global-symmetry violating effects to salvage
axions as a solution to the strong CP problem (which is a much
greater suppression than that necessary to satisfy the bounds
on the $\beta_i$'s above).

Evidently it is hard to estimate reliably the degree to which 
an approximate global symmetry can consistently be invoked in a 
world with gravity.  Given the tentative character of our current 
understanding, we should take seriously
the possibility that the quintessence field has avoided direct
detection because the couplings considered above are suppressed
by such a symmetry.

An important consequence of this viewpoint
is that interactions which are invariant
under $\phi \rightarrow \phi+{\it const}$ --- that is,
derivative couplings of $\phi$ --- should be present with 
couplings $\beta_i$ of order unity.  The derivative term of lowest
dimension that could multiply an arbitrary gauge-invariant scalar
operator would be $g^{\mu\nu}\nabla_\mu\nabla_\nu\phi$; however,
we would expect this dimension-three term to be divided by $M^3$
and hence lead to negligible effects. The other possibility is
to couple $\phi/M$ to a total derivative, which after integration
by parts is equivalent to a coupling to $\partial_\mu\phi$.  The only
allowed term in the standard model is
\be
  \beta_{F\widetilde F}{\phi\over M} F_{\mu\nu}\widetilde F^{\mu\nu}
  = {{\beta_{F\widetilde F}}\over{M}}\left[-(\partial_\mu\phi)K^\mu
  + \partial_\mu(\phi K^\mu)\right]
  \ ,
  \label{ffdual}
\ee
where $F_{\mu\nu}$ is the electromagnetic field strength tensor,
$\widetilde F^{\mu\nu} = {1\over 2}\epsilon^{\mu\nu\rho\sigma}
F_{\rho\sigma}$ is its dual, and $K^\mu = 2A_\nu \widetilde F^{\mu\nu}$.
The divergence term on the right hand side of (\ref{ffdual}) 
contributes a surface term to the action, which vanishes for fields
which fall off at infinity.  Therefore this interaction does represent
a derivative coupling, and respects the symmetry $\phi \rightarrow
\phi + {\it const}$.   

Such a coupling can lead to potentially observable effects.
Since $F_{\mu\nu}\widetilde F^{\mu\nu}$ is a pseudoscalar quantity,
it does not accumulate coherently in a macroscopic test body, and
hence does not give rise to appreciable long-range forces (although
one can consider tests using polarized bodies \cite{poltest}).
However, 
a spatially homogeneous but slowly-varying $\phi$ field would rotate
the direction of polarization of light from distant
radio sources \cite{cfj}.  The dispersion relation for electromagnetic
radiation in the presence of a time-dependent $\phi$
becomes
$\omega^2 = k^2 \pm (\beta_{F\widetilde F}/M)\dot\phi k$,
where $+/-$ refer to right- and left-handed circularly polarized
modes, respectively.  If we define $\chi$ to be the angle between
some fiducial direction in the plane of the sky and the polarization
vector from an astrophysical source, then in the WKB limit where
the wavelength of the radiation is much less than that of $\phi$,
the difference in group velocity for the two modes leads to a 
rotation $\Delta\chi = (\beta_{F\widetilde F}/M)\Delta\phi$.

Such a rotation is potentially observable, as distant radio galaxies
and quasars often have a well-defined relationship between their
luminosity structure and polarization structure \cite{galaxies}.  In 
the wake of a claim that a dipole pattern of rotations (in contrast to the
monopole pattern expected from a homogeneous scalar field) was present 
in existing data \cite{nr}, it was pointed out that more recent observations 
provide a stringent upper limit on any such effect \cite{leahy,wpc}.  It 
is a straightforward exercise to use these same data to place upper limits
on the magnitude of a direction-independent pattern of 
rotations.  As an example, Figure One shows the data given by
Leahy \cite{leahy} for $\Delta\chi$, plotted as a function of redshift.
   \begin{figure}
   \centerline{
   \epsfxsize=3.3 in \epsfbox{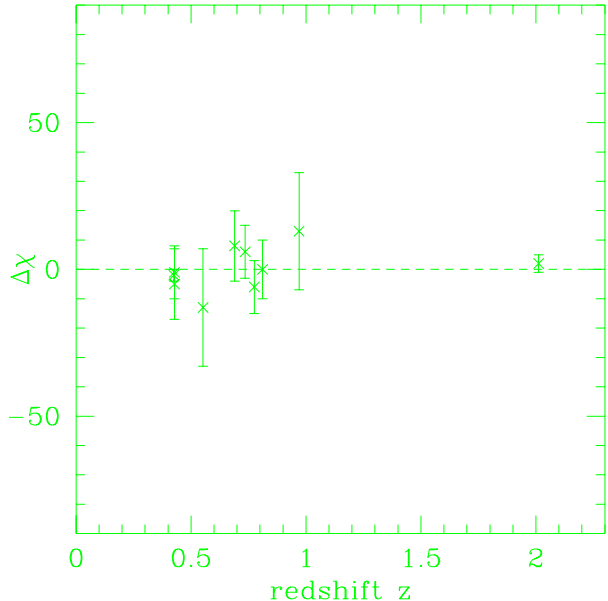}
   }
   \caption{Rotation of polarization vector vs. redshift.}
   \end{figure}
Simply taking the mean value all of the points (for which the minimum
redshift is $z=0.425$) yields $\langle\Delta\chi\rangle = -0.6^\circ 
\pm 1.5^\circ$. This implies a bound
\be
  |\beta_{F\widetilde F}| \leq 3\times 10^{-2} \left({M\over
  {|\Delta\phi|}}\right)\ ,
  \label{radiolimit}
\ee
where $\Delta\phi$ is the change in $\phi$ between a redshift 
$z=0.425$ and today.
From the figure, it is evident that the single source 3C~9 at a 
redshift $z=2.012$ (originally analyzed in \cite{kdr}),
with $\Delta\chi = 2^\circ \pm 3^\circ$, provides
an interesting limit on any substantial rotation at high redshifts.

Does Eq.~(\ref{radiolimit}) constitute a good limit?  We expect
$\Delta\phi$ to be of order $\mp$, so the factor $M/|\Delta\phi|$
is likely to be less than or of order unity.  However, while
$3\times 10^{-2}$ is less than one, it is not remarkably less; we
might imagine that dimensionless constants conspire to make 
$\beta_{F\widetilde F}$ naturally smaller than this value even without
suppression by some symmetry.  For example, couplings of the form
(\ref{ffdual}) can arise through triangle graphs in ordinary
field theories ({\it i.e.}, even disregarding the possibility of
exotic quantum gravitational effects); such graphs lead to 
$\beta_{F\widetilde F} = N\alpha/4\pi$, where $\alpha$ is the
fine structure constant and $N$ is a dimensionless factor which
depends on the field content of the model.  
Since $\alpha/4\pi\sim 6\times 10^{-4}$,
it is by no means implausible that the interaction under consideration
could exist but have evaded detection thus far.  This raises the
exciting possibility that improvements in the limits from radio
galaxy polarization measurements could lead to a detection of
quintessence.  Since the relevant observed quantity is an angle,
it is hard to imagine significant contamination by systematic
errors, so the observation of a large number of sources can
be expected to improve these limits substantially.

Unfortunately, the existence of a potentially detectable coupling of the
form (\ref{ffdual}) can be avoided in certain models.  This
follows from noting that the analogous term for the
strong interactions, $\beta_{G\widetilde G}(\phi/M)\tr(G_{\mu\nu}
\widetilde G^{\mu\nu})$, is not invariant under $\phi \rightarrow
\phi + {\it const}$ due to the existence of topologically
nontrivial field configurations.  The surface term which could
be neglected in electromagnetism would receive contributions from
QCD instantons, leading to a mass for $\phi$ proportional to
$\beta_{G\widetilde G} (\Lambda_{\rm QCD}^2/M)$ (just as for the QCD 
axion).  As this mass is likely to be much larger than the desired value
$m_\phi\sim 10^{-33}$~eV, it is incompatible with the desired
properties of quintessence.  In a grand unified model for which
both electromagnetism and the strong interactions derive from a
single simple gauge group, any gauge-singlet field which couples to
$F_{\mu\nu}\widetilde F^{\mu\nu}$ should also couple to $\tr(G_{\mu\nu}
\widetilde G^{\mu\nu})$ \cite{axcoup}.  In the minimal SU(5) grand unified
theory, for example, the appropriate lowest-dimension
gauge-invariant operator to which $\phi$
could couple is $\tr(V_{\mu\nu}\widetilde V^{\mu\nu})$, where
$V_{\mu\nu}$ is the SU(5) field strength.  After spontaneous
symmetry breakdown this term includes a unique linear combination
$\tr(G_{\mu\nu}\widetilde G^{\mu\nu}) + {4\over 3}F_{\mu\nu}
\widetilde F^{\mu\nu}$ with which $\phi$ could interact.  Since
the coupling to the QCD term must be suppressed, the electromagnetic
coupling will be suppressed as well.

This argument undoubtedly diminishes the aura of inevitability
surrounding a coupling of the form (\ref{ffdual}), but by no means
precludes its existence.  A simple way out is to 
imagine that SU(3) and U(1) are not unified in a simple gauge
group, in which case there is no necessary relationship between
the QCD and electromagnetic couplings.  Such a scenario may be
natural in string theory, where low-energy gauge fields come from
compactification as well as the original gauge symmetry in
higher dimensions.
Another way is to include
a coupling of $\phi$ to higher-dimensional gauge-invariant
operators through interactions such as 
$(\phi/M^2)\tr(\Sigma V_{\mu\nu}\widetilde V^{\mu\nu})$,
where $\Sigma$ is the adjoint Higgs which breaks SU(5).  
If the mass scale $v$ is
comparable to $M$, such an interaction could cancel the QCD
term from $\tr(V_{\mu\nu}\widetilde V^{\mu\nu})$, leaving an
unsuppressed coupling to electromagnetism.  
However, the interaction 
$(\phi/M^2)\tr(\Sigma V_{\mu\nu}\widetilde V^{\mu\nu})$
is not invariant under $\phi \rightarrow \phi + const$, so
it may be noticeably suppressed.

In conclusion, the absence of observable interactions of
quintessence with the fields of the standard model implies
the existence of a symmetry which suppresses such couplings.
Such a symmetry leaves open the possibility of a coupling to
electromagnetism, which is potentially observable in polarization
studies of distant radio sources.  Such a coupling is not inevitable,
however, so we may have to rely on conventional cosmological
tests to determine whether slowly rolling scalar fields 
play an important role in the dynamics of the present universe.

\section*{Acknowledgments}

I have benefited from helpful comments from numerous people.
I would like to especially thank Tom Banks, Chris Hill, Joe
Polchinski, and Mark Srednicki for extensive conversations.
This research was supported by the National Science Foundation
under grant PHY/94-07194.

\vfill

\end{document}